\documentclass[aps,prb,twocolumn,showpacs,groupedaddress,amsmath,amssymb]{revtex4}

\usepackage{ulem}
\usepackage{graphicx}
\usepackage{color}
\usepackage{comment}
\usepackage{bm}
\usepackage{latexsym}
\usepackage[latin9]{inputenc}
\usepackage{amsmath}
\usepackage{esint}
\usepackage{braket}

\usepackage[english]{babel}

\usepackage[symbol]{footmisc}
\usepackage{hyperref}

\usepackage[T1]{fontenc}
\usepackage[latin9]{inputenc} 

\begin{document}

\title{Optical Phase Transitions in Photonic Networks: A Spin System Formulation}
\author{Alba Ramos, Tsampikos Kottos\footnote{corresponding author:tkottos@wesleyan.edu}} 
\affiliation{Wave Transport in Complex Systems Lab, Department of Physics, Wesleyan University, Middletown, CT-06459, USA}
\author{Boris Shapiro} 
\affiliation{Technion - Israel Institute of Technology, Technion City, Haifa 32000, Israel}
\date{\today}

\begin{abstract}
We investigate the collective dynamics of nonlinearly interacting modes in multimode photonic settings with long-range couplings. To 
this end, we have established a connection with the theory of spin networks. The emerging ``photonic spins'' are complex, soft (their 
size is not fixed) and their dynamics has two constants of motion. Our analysis shed light on the nature of the thermal 
equilibrium states and reveals the existence of optical phase-transitions which resemble a paramagnetic to a ferromagnetic and to a 
spin-glass phase transitions occurring in spin networks. We show that, for fixed average optical power, these transitions are driven 
by the type (constant or random couplings) of the network connectivity and by the total energy of the optical signal. 
\end{abstract}

\maketitle

\section{Introduction}
In physics one often encounters problems involving a great number of nonlinear interacting modes. Such problems naturally arise in 
statistical mechanics \cite{FPU74,PB11,J57}, hydrodynamics \cite{ZLF12,NR12}, matter-waves \cite{TCFMSEB12,SFSGPK18,CKGM15}, 
and more. An emerging framework is in photonics, where light propagation in non-linear multimode optical structures have recently attracted 
a lot of attention \cite{SLBSM09,KSVW10,SPS11,SJBRPF12,RW13,SMF13,Petal14,WCW15,KTSFBMWC17,WCW17,WHC19}. On 
the fundamental side there are many unanswered questions associated with the energy exchange between the modes and the role of 
the underlying spatio-temporal complexity, originating from the disorder, the network topology and the complex intermodal interactions. 
Brute-force computational attempts to answer these questions are either impossible (due to the large number of degrees of freedom 
involved) or unsatisfactory as far as the understanding of the underlying physics that dictates the energy redistribution. At the same time, 
there is a pressing need from modern technologies to develop theoretical tools that will allow us to tailor the intermodal energy exchange 
and harvest it to our advantage. If this endeavor is successful, it will give rise to a next generation of high power light sources \cite{WCW17}, 
high-resolution imaging schemes \cite{BWF09,MLLF12,PTC15}, and high-speed telecommunication systems \cite{RFN13,FK05,HK13}.

\begin{figure}
\includegraphics[width=\columnwidth]{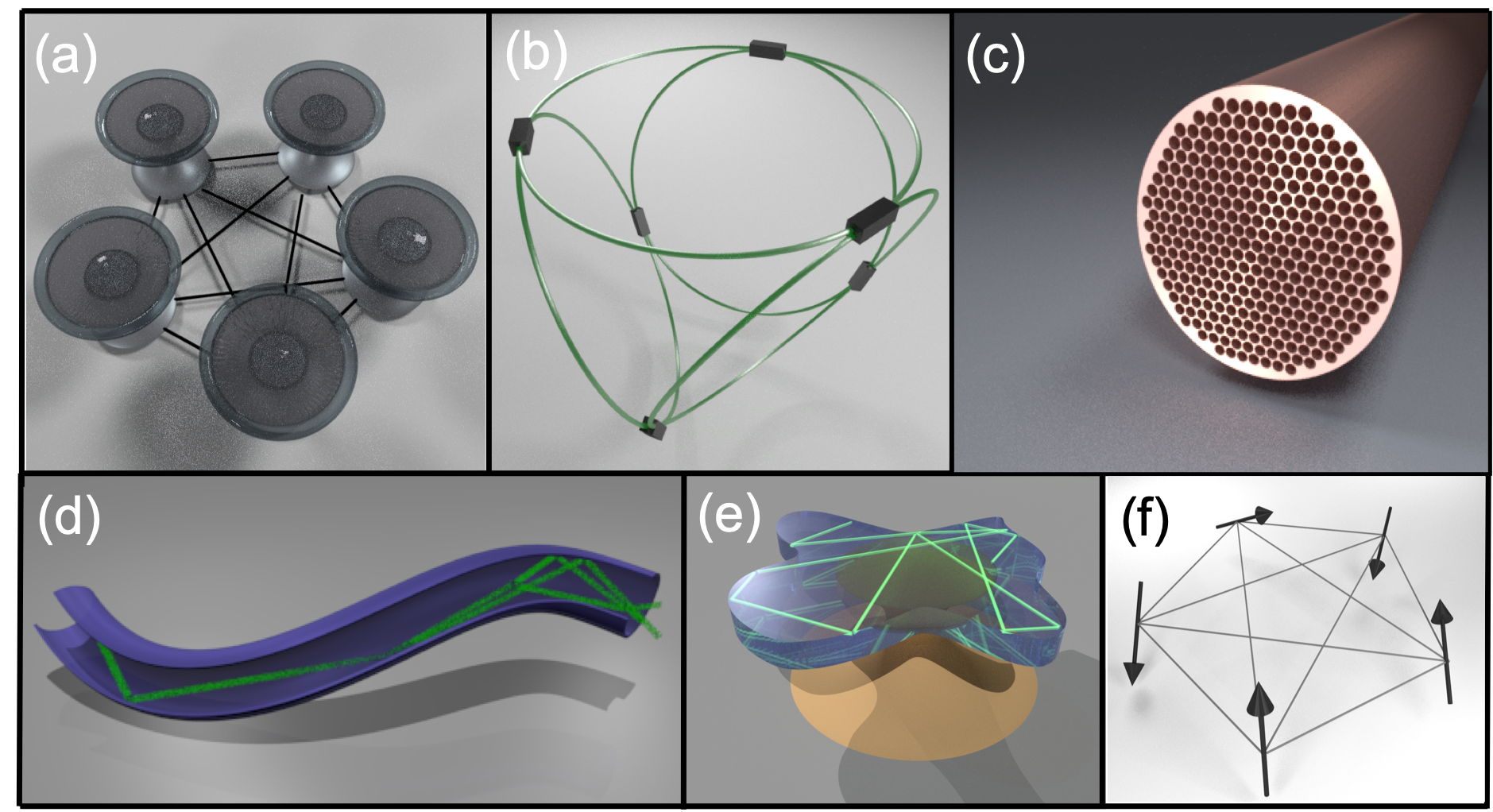}
\caption{(Color online) A variety of photonic nonlinear multimode networks, with long-range modal couplings, whose field dynamics is 
modeled by an effective coupled mode theory of the type given by Eqs. (\ref{dif_schr},\ref{dif_schr_mode}). (a) A network of coupled 
micro-resonators \cite{V03}; (b) A photonic network of single-mode fibers coupled together via couplers (optical splitters) \cite{LTG17,
GSBPCS19}; (c) A multicore fiber \cite{HK13};  (d) A multimode fiber \cite{HK13,CLXHCK18,LCK19}; (e) Deformed multimode (micro-)
resonator with underlying chaotic dynamics \cite{FS97,BCH00,HSKBS01}; (f) A network of coupled ``soft'' (size-modulated) spins. The 
corresponding Hamiltonian has similarities but also crucial differences with the Hamiltonian that describes a photonic multimode network.
\label{fig1}}
\end{figure} 

In this regard, a number of recent papers have promoted well established equilibrium \cite{SLBSM09,SPS11,WHC19,PWJMC19,BFS17,KS11} 
and non-equilibrium \cite{Petal14,DNPZ92,P07,AGMDP11,PR12} thermodynamics techniques as a viable theoretical toolkit which can 
be used to address some of the above challenges. A decisive step along these lines has been achieved recently by the authors of Ref. 
\cite{WHC19,PWJMC19} which, assuming thermal equilibrium conditions and {\it weak} non-linearity, have established a comprehensive 
optical thermodynamics formalism allowing us to design potentially novel classes of high-power multimode optical structures or efficient 
cooling schemes. For weak nonlinearity, the specific nature of the nonlinear mode interaction (e.g. Kerr or saturable or thermal nonlinearities) 
is irrelevant. Its role is important for the thermalization process but not for the properties of the equilibrium state.

The implementation of statistical thermodynamics methods in modern photonics opens up a new arena where ideas and concepts from 
statistical mechanics can be transfered to optics and utilized for light control. A prominent example is the notion of phase-transitions 
characterized by changes in the properties of a system as an external parameter varies \cite{PB11}. Maybe, the most celebrated 
area of physics where such phenomena have been extensively studied is the theory of spin network models \cite{s1,s2,s3}. These 
studies indicated that the phase transition can be understood as a competition between the interactions among spins, which facilitates 
order, and the thermal fluctuations (the entropy) causing random disturbances. In many occasions, these phase transitions are 
associated with symmetry breaking phenomena. To understand them, one has to analyze the nature of the thermal equilibrium state of 
the interacting system. In this paper, we will show that there are certain analogies between the statistical thermodynamics of spin models 
and multimode photonic networks. Inspired by these analogies we will ask questions like: are the phases of the electromagnetic 
field correlated or are they entirely random? Is the optical power distributed more or less evenly between all modes of the entire system, 
or some finite fraction of the total optical power can reside in a single mode? How the topology of the inter-mode connectivity and the 
randomness in the coupling affect the nature of the thermal state?

It turns out that these questions can be related. In previous works, for example, it was shown that macroscopic occupation of a single 
photonic mode does occur and, moreover, it can happen even in linear systems (we stress again that weak non-linearity can be neglected 
only in the equilibrium state, while it is crucial in the thermalization process) \cite{BFS17,AGMDP11,note1}. It is quite remarkable, thus, 
that a purely classical system exhibits a phenomenon alike BEC transition in a quantum Bose gas. Actually, it has been a number of 
experiments demonstrating that, in the course of propagation along the fiber, optical power ``flows'' towards the lower modes 
\cite{GEEAWWC2016,LWCW16,KTSFBMWC17,N19}. 

In this paper, we demonstrate that the connectivity of an optical structure is an important factor in its thermalization process: it affects 
the type of the optical phase transitions and the nature of the thermal equilibrium state. The question is not only of fundamental importance;
it pertains also to recent photonic developments where networks with complex connectivities, can be realized, see Figs. \ref{fig1}a-e. 
Specifically, we show that in the case of long-range couplings, the nonlinearity is instrumental for achieving optical phase transitions. 
In fact, by identifying an order parameter that is equivalent to the magnetization in spin-network models, we are able to show both 
theoretically and numerically the existence of a ferromagnetic-to-paramagnetic phase transition, analogous to the one occurring in spin 
systems. When disorder is introduced into the couplings, the system might undergo another type of a transition, namely to a spin-glass 
phase; much as in the case of frustrated coupled spins. Although these analogies between photonics multimode networks (Figs. \ref{fig1}
a-e) and spin-networks (Fig. \ref{fig1}f) is useful, one needs to keep in mind that the two problems have important differences. 
Specifically our ``photonic spins'' are complex dynamical variables (amplitudes of the electric field). Moreover, they fluctuate not only 
in their direction but also in their size. We expect that the analogies drawn from our study will bring together two seemingly different 
areas: statistical mechanics of spin networks and light transport in nonlinear multimode settings. This cross-fertilization will, 
hopefully, allow the development of better design strategies for the control of light transport in multimode photonic networks.
 
The structure of this paper is as follows. In the next section \ref{genform} we discuss the general statistical thermodynamics formalism
associated with the analysis of optical thermal equilibrium states. Special attention is given to the case of weak nonlinearities where we 
derive the occupation number statistics. In the next section \ref{lrc} we analyze a class of complex multimode photonic networks with 
long-range connectivity. Two cases are discussed in detail: the case of constant couplings and the opposite case of random couplings. 
We show that under specific conditions these systems demonstrate optical phase transitions from ferromagnetic to paramagnetic and 
spin-glass phases. Finally, our conclusions are discussed at the last section \ref{concl}.

\section{General Formalism}\label{genform}

The dynamics of  nonlinear multimode photonic networks in Fig. \ref{fig1} can be modeled using the framework of time dependent coupled 
mode theory. The associated equations are
\begin{equation}
 i\frac{d\psi_l}{dt}=-\sum_j J_{lj}\psi_j+\chi\left|\psi_l\right|^2\psi_l,\quad l=1,\cdots,N
 \label{dif_schr}
 \end{equation}   
where $\psi_l$ is the degree of freedom (the complex amplitude) at node $l$ of the ``network'', and $J_{lj}=J^*_{jl}$ is the connectivity matrix 
that dictates the couplings among the nodes. We will typically assume zero self-coupling terms $J_{ll}=0$. Finally, the last term in Eq. (\ref{dif_schr}) 
describes the nonlinearity due, for instance, to Kerr effect.  

On a formal and general level $\psi_l=\langle l |\psi\rangle$ are components of the electric field (with some fixed polarization) in some basis of 
orthonormal modes $\{\left| l\rangle\right.\}$ (the ``basic modes''). The choice of the set of these modes depends on the problem at hand. For 
instance, for the case of coupled single-mode microresonators \cite{V03} (Fig.\ref{fig1}a) the index $l$ labels the resonators and the ``basic 
mode'' $\left| 
l\rangle\right.$ is the eigenmode of the $l$-th resonator, decoupled from the rest of the network (we treat the resonators as structureless point 
objects). In this framework, the coefficients $J_{lj}=J^*_{jl}$ (for $l\neq j$) represent evanescent couplings between different resonators and 
$J_{ll}=\omega_l$ is the eigenfrequency of resonator $l$ (with nonlinearity neglected), which in case of identical resonators can be set to be 
zero i.e. $\omega_l=0 (l=1,\cdots,N$). The same interpretation applies to a fiber network \cite{LTG17,GSBPCS19} Fig. \ref{fig1}b and to a multicore 
fiber \cite{HK13}, Fig. \ref{fig1}c, where now $l$ labels the single mode fibers. In Fig. \ref{fig1}c the propagation direction $z$ plays the role of 
time.

It is important to clearly distinguish between the ``basic modes'', like an eigenmode of an isolated resonator in Fig.\ref{fig1}a, and the eigenmodes 
of the entire structure, i.e. the stationary solutions of the entire system of coupled equations (\ref{dif_schr}) (with the nonlinearity neglected). The 
latter are often referred to as ``supermodes''. For instance, when we are talking about condensation of the optical power in a single mode, we mean 
of course the supermode (with the lowest energy) and not an eigenmode of a single resonator. Sometimes, where no confusion can arise, we will 
use the term ``mode'' instead of ``supermode''. The ``supermodes'' $f_{\alpha}(l)$ form a complete basis, so the field $\psi_l(t)$ can be expanded as 
$\psi_l(t)=\sum_\alpha C_\alpha(t)f_\alpha(l)$, reducing Eq. (\ref{dif_schr}) to
\begin{equation}
 i\frac{dC_\alpha}{dt}=\varepsilon_\alpha C_\alpha(t)+\chi\sum_{\beta\gamma\delta}\Gamma_{\alpha\beta\gamma\delta}C_\beta^*(t)
C_\gamma(t)C_\delta(t),
\label{dif_schr_mode}
\end{equation}
with
\begin{equation}
 \Gamma_{\alpha\beta\gamma\delta}=\sum_lf_\alpha^*(l)f_\beta^*(l)f_\gamma(l)f_\delta(l).
\end{equation}
As we will see below, this mode representation of the evolution Eq. (\ref{dif_schr}) is particularly useful when the non-linearity is weak.

Equations (\ref{dif_schr}) and (\ref{dif_schr_mode}) also describe multimode optical fibers \cite{HK13,CLXHCK18,LCK19} or multimode resonators 
\cite{FS97,BCH00,HSKBS01}, Figs. \ref{fig1}d,e respectively. In this case the ``basic modes'' $\left| l\rangle\right.$ are 
the eigenmodes of an ideal, undeformed fiber (or resonator) while $J_{lj}$ are the couplings among these modes, due to various perturbations 
(deformations of the ideal system, possible impurities, etc). Note that unlike the previous case , when the basic modes were localized in space 
(on a single resonator), now they extend over the entire structure.

The equation of motion (\ref{dif_schr}) is derivable from the energy functional (the Hamiltonian)
\begin{equation}
 {\cal H}\{\psi_l(t)\}=-\sum_{l,j}J_{l,j}\psi^*_l\psi_j+\frac12 \chi\sum_l\left|\psi_l\right|^4\equiv E.
 \label{Hamiltonian}
\end{equation}
In the course of time the total energy $E$ and the total optical power
\begin{equation}
 {\cal N}\{\psi_l(t)\}=\sum_l\left|\psi_l(t)\right|^2\equiv A,
\label{norm}
\end{equation}
are conserved. Finally, we always assume that both the total power $A$, and the energy $E$, are extensive quantities, proportional to 
the number of modes $N$, as appropriate for thermodynamics.  
\subsection{The Problem of Thermalization}

An important question is whether an isolated system of interacting modes eventually thermalizes, i.e. reaches an equilibrium state 
which can be described by just two parameters -the inverse temperature $\beta$ and the chemical potential $\mu$, which in turn 
are determined by the energy $E$ and the total power $A$ of the initial preparation. If such equilibrium state is reached, then the 
system can be analyzed using the well established methods of statistical mechanics and thermodynamics. 

For example, a statistical mechanics description of the system of Eqs. (\ref{dif_schr}) is achieved by calculating the classical grand
-canonical partition function ${\cal Z}$
\begin{equation}
 {\cal Z}=\int\left(\prod_{l=1}^Nd\psi_l^*d\psi_l\right)e^{-\beta{\cal H}+\beta\mu{\cal N}}
\label{Zfunction}
\end{equation}
where the Lagrange multipliers $\beta=1/T$ and $\mu$ have been introduced (in analogy with the inverse temperature and the 
chemical potential) to ensure conservation (on average) of $E$ and $A$ respectively (see Eqs. (\ref{Hamiltonian},\ref{norm})). 
Specifically the relation between the microcanonical quantities
\begin{equation}
a\equiv\frac{A}{N}, \qquad h\equiv\frac{E}{N}
\label{aE}
\end{equation}
which describe the average optical power $a$ and averaged energy density $h$ per mode and the grant canonical quantities 
$\mu,\beta$ is given by
\begin{equation}
 a=\frac{1}{\beta N}\frac{\partial ln({\cal Z})}{\partial \mu};\quad
 h=-\frac1N \frac{\partial ln({\cal Z})}{\partial\beta}+\frac{\mu}{N\beta}
 \frac{\partial ln({\cal Z})}{\partial\mu}.
\label{a_h}
\end{equation}

Using the partition function as a starting point we can next calculate the thermodynamic potential
\begin{equation}
\label{omega0}
 \Omega=-T\,ln\left({\cal Z}\right)
\end{equation}
and from there, the entire ``optical thermodynamics'' can be derived. For instance, the entropy is $S=-\left(\frac{\partial\Omega}
{\partial T}\right)_\mu$.

Finally we note that the problem of thermalization in non-linear lattices, under time evolution defined in Eq. (\ref{dif_schr}) (primary 
in one spatial dimension and with $J_{lj}$ restricted to nearest neighbors only) has been also addressed in the framework of 
statistical mechanics \cite{RCKG00,JR04,R08,LS18}. In this studies, it has been pointed out that thermalization occurs only in a 
certain region of the $(E,A)$-plane. These studies indicated that for fixed total norm $A$, the system thermalizes only if its energy 
is not too large \cite{RCKG00}. Otherwise, the equilibrium Gibbs distribution, Eq. (\ref{Zfunction}), cannot be reached- the system 
is said to belong to the non-Gibbsian, or negative temperature region. The maximum value $h=h_{max}$ for a given average norm 
$a$ per site corresponds to $\beta\longrightarrow0$ (high temperature) and $\mu\longrightarrow-\infty$, and it is
\begin{equation}
\label{eq_hmax}
 h_{max}=\chi a^2,
\end{equation}
It turns out that for sufficiently large $E$ the norm cannot spread uniformly over the entire system and high-amplitude peaks of 
$\psi_l$ (breathers \cite{FG08}) emerge. Below, we will confine our analysis to the domain where the temperature is positive 
and thermalization can be achieved.

\subsection{Thermal Equilibrium in the Case of Weak Non-linearities}

In many optics applications the non-linearity is considered weak. It is of course essential in the mode-mixing process (see Eq. 
(\ref{dif_schr_mode})), needed for reaching equilibrium. The total energy and power, however, are dominated by the linear 
term in the Hamiltonian, i.e. to a good approximation
\begin{equation}
 E=\sum_{\alpha=1}^N \varepsilon_\alpha|C_\alpha|^2, \qquad A=\sum_{\alpha=1}^N |C_\alpha|^2.
 \label{constrai}
\end{equation}
Here $|C_\alpha|^2$ is the (normalized) optical power in mode $\alpha$. Using these expressions Christodoulides {\it et al.} 
\cite{WHC19,PWJMC19} were able to develop a kind of ``optical thermodynamics'', identifying the optical analogy of entropy, 
equation of states and other quantities. Below we briefly summarize this theory, using the grand-canonical formulation Eq. 
(\ref{Zfunction}) discussed above. 

Assuming that the system has thermalized, its grand partition function, non-linearity being neglected, is given by the product of 
independent modes contributions
\begin{equation}
{\cal Z}=\prod_{\alpha=1}^N\left[\int dC_\alpha^*dC_\alpha e^{-\beta\left(\varepsilon_\alpha-\mu\right)|C_\alpha|^2}\right]=
\prod_{\alpha=1}^N\left[\frac{\pi}{\beta\left(\varepsilon_\alpha-\mu\right)}\right],
\label{grand_can}
\end{equation}
where $\beta=1/T$ and $\mu$ can be found from the constraints in Eq. (\ref{constrai}). It immediately follows from Eq. (\ref{grand_can}) 
that the average of optical power in mode $\alpha$ obeys the Rayleigh-Jeans distribution \cite{WHC19, PWJMC19,DNPZ92,P07,
AGMDP11,BFS17} (see also Ref.\cite{ES13})
\begin{equation}
 \left\langle |C_\alpha|^2\right\rangle=\frac{T}{\varepsilon_\alpha-\mu}\equiv \bar{n}_\alpha.
 \label{C_alpha}
\end{equation}
From Eq. (\ref{C_alpha}) we can calculate the thermodynamic potential Eq. (\ref{omega0}) which takes the form
\begin{equation}
\label{omega}
 \Omega=-T\,\sum_{\alpha=1}^N\ln\left(\frac{\pi T}{\varepsilon_\alpha-\mu}\right).
\end{equation}
All other thermodynamic variables follow from Eq. (\ref{omega}). For example, the entropy (up to an irrelevant constant) is $S=
\sum_{\alpha=1}^N\ln\left(\bar{n}_\alpha\right)$ which, in equilibrium, coincides with the expression derived in Ref. \cite{WHC19} 
by counting the number of ways in which a large number of ``packets of power'' can be distributed over the $N$ modes. The 
expression for $S$ served in \cite{WHC19} as the starting point for the development of ``optical thermodynamics''. For instance, 
one can derive the following equation of state\cite{WHC19} $E-\mu A=NT$ which connects three extensive quantities $(E,A,N)$ 
to the two intensive variables $(\mu, T)$. 
\subsection{Fluctuations in the Case of Weak Non-linearities}

Next, we briefly discuss the fluctuations of the optical power $n_{\alpha}$ in the mode $\alpha$. If the nonlinearity, i.e. the intermode 
interaction, in the thermal equilibrium state is negligibly small \cite{WHC19,PWJMC19,DNPZ92,P07,AGMDP11,BFS17}, then within 
the grand canonical treatment the probability density for $n_{\alpha}$ is ${\cal P}(n_{\alpha})={1\over {\cal Z}_{\alpha}}\exp\left[-\beta
\left(\varepsilon_{\alpha}-\mu\right)n_{\alpha}\right]$, where ${\cal Z}_{\alpha}$ is the normalization factor. This yields Eq. (\ref{C_alpha}) 
for the average value ${\bar n}_{\alpha}$ and $\overline{\Delta n_{\alpha}^2}=\left({\bar n}_{\alpha}\right)^2$ for the variance. The 
same results can be obtained, in even simpler way, if one uses the expression $\Omega_{\alpha}=-T\ln{\pi T\over \varepsilon_{\alpha}
-\mu}$ for the contribution of mode $\alpha$ to the grand potential $\Omega$ (see Eq. (\ref{omega})) and the standard formulas 
\cite{LLv1} ${\bar n}_{\alpha}=-\partial\Omega_{\alpha}/\partial\mu$,  $\overline{\Delta n_{\alpha}^2}=T\partial {\bar n}_{\alpha}/\partial 
\mu$.

Thus, the standard deviation $\left(\overline{\Delta n_{\alpha}^2}\right)^{1/2}\equiv \sigma_{\alpha}$ comes out to be equal to the 
average optical power $\bar{n}_{\alpha}$. For a mode with macroscopic occupation $(n_{\alpha}\gg 1)$ this result looks paradoxical.
The ``paradox'', however, is well known in the theory of Bose-Einstein condensation and it is resolved by observing that we have here 
one of the rare cases when the canonical and the grand canonical ensembles yield different results \cite{LLv1}. Indeed, in the 
experiment, as well as in our numerical simulations, the total optical power $\sum_{\alpha}n_{\alpha}=A$ is strictly conserved while in 
the grand canonical treatment it is conserved only on the average. This is perfectly fine for calculating various average quantities but 
not for the fluctuations. When the conservation law $\sum_{\alpha}n_{\alpha}=A$ is strictly enforced (canonical ensemble), the large
unphysical fluctuations in a macroscopically occupied mode disappear (for instance, at $T=0$, when the entire power $A$ is located 
on a single mode, there are no fluctuations at all).

Note, however, that at high temperatures, when there are many modes populated with $\bar{n}_{\alpha}\lesssim 1$, the result $\sigma_{\alpha}
=\bar{n}_{\alpha}$ does hold for such modes. This is because the constraint $\sum_{\alpha}n_{\alpha}=A\sim N$ on the total power 
does not significantly affect fluctuations in a single mode with $\bar{n}_{\alpha}\lesssim 1$ (the other modes serve as an ``environment" 
for the mode $\alpha$). One should be aware of these large fluctuations when interpreting the numerical or the experimental data.

\section{Multimode Optical Systems with Long Range Coupling}\label{lrc} 

We consider the connectivity matrix $J_{lj}$ (see Eq. (\ref{dif_schr})) of the following form:
\begin{equation}
\label{conmat}
J_{lj} = \left({J_0\over N} + {\sigma\over \sqrt{N}}B_{lj}\right)(1-\delta_{lj}),     
\end{equation}
where the first term describes a fully connected network, with equal couplings, while the second term introduces some randomness 
into the couplings. In the case of chaotic or disordered networks \cite{stoe07} the couplings $B_{lj}=B_{jl}$ are given by a Gaussian 
distribution with zero mean and a unit standard deviation. We point out that, unless stated otherwise, in all simulations below the 
random matrix elements $B_{i,j}$ remain the same (fixed) for a specific set of parameters $N,J_0,\chi$ dictating the Hamiltonian of 
the photonic network. For ${\cal H}$ in Eq. (\ref{Hamiltonian}) to remain extensive, in the large $N$ limit, one has to scale the coupling 
strengths with $N$, as written in Eq. (\ref{conmat}).

The coupling matrix Eq. (\ref{conmat}) gives rise to two distinct terms in the Hamiltonian Eq. (\ref{Hamiltonian}). The first term resembles 
the Curie-Weiss (CW) model \cite{FV17}, where $N$ Ising spins are coupled to each other by constant distance-independent interactions. 
The second term 
resembles the Sherrington-Kirkpatrick (SK) model for a spin glass \cite{SK75,EA75}, where the couplings are completely random. We write 
``resembles'' because our model differs from the well studied CW and SK models in three important respects: First, our dynamical variables 
${\psi_l}$ are complex, as appropriate for the complex amplitudes of electric field, and they can be treated as real two-component ``spins''. 
Second, since $|\psi_l|^2$ is not restricted to some fixed value, our spins fluctuate not only in their direction but also in their size (note though, 
that the non-linearity does not allow for too wild fluctuations in size). And, third, the dynamics of our ``photonic spins'' conserve not only the
energy (as in standard spin systems) but also the optical power, see  Eq. (\ref{norm}). This second conservation law introduces novel features
into the characteristics of the thermal equilibrium state; for instance a condensation of optical power in a single mode. 

Below we will distinguish between the two limiting cases corresponding to a photonic network with equal couplings ($\sigma=0$) and to its 
``random'' coupling analogue ($J_0=0$). We will also briefly discuss the case where the connectivity matrix Eq. (\ref{conmat}) contains both 
terms. We will show that our photonic network exhibits various phases depending on the disorder strength of the coupling constants, the energy 
and optical power $(h,a)$ of the initial preparation and the strength of the nonlinearity parameter $\chi$.

\subsection{Numerical Method}

The thermalization process of an initial state $\{\psi_n\} (n=1,\cdots,N)$ has been investigated numerically using a high order three 
part split symplectic integrator scheme \cite{Skokos2014,Skokos2016,Channell-Neri} for the integration of Eq. (\ref{dif_schr}). The 
method conserved, up to errors ${\cal O}(10^{-8})$, the total energy Eq. (\ref{Hamiltonian}) and the optical power Eq. (\ref{norm}) of 
the system. These quantities have been monitored during the simulations in order to ensure the accuracy of our results.

We have focused our interest on the electric field amplitudes $\psi_n(t)$ and the supermode amplitudes $C_\alpha(t)$ which can 
be evaluated from the projection of $\psi_n(t)$ on the supermode basis, see Eq. (\ref{dif_schr_mode}). Knowledge of $C_\alpha(t)$ 
(or $\psi_n(t)$) allows us to calculate various thermodynamic quantities $\langle Q\rangle$ by making a time-average and invoking 
ergodicity.

\begin{figure}
\includegraphics[width=\columnwidth]{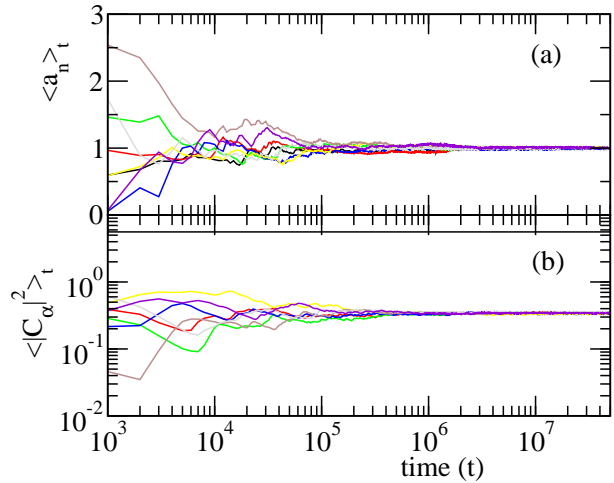}
\caption{(Color online) An example (a) of a time-averaged nodal power $<a_n>_t={1\over (t-t_{\rm min})}\int_{t_{\rm min}}^t |\psi_n(t)|^2 dt$  
and (b) of a supermode power $<|C_{\alpha}|^2>_t={1\over (t-t_{\rm min})}\int_{t_{\rm min}}^t |C_{\alpha}(t)|^2 dt$. In these simulations 
the photonic network consists of $N=8$ nodes, $J_0=1.2$ and $\sigma=0$. The initial preparation has averaged energy density $h\approx 
-0.61$ and optical power $a=1$. The nonlinearity parameter is $\chi=0.01$. The initial time is $t_{\rm min}=1000$ in both cases. 
\label{fig2}}
\end{figure} 
 
In practice, the approach to a thermal equilibrium state involves a long time propagation of an initial preparation $\{\psi_n\} (n=1,\cdots,N)$. 
Typical integration times were as long as $4\times 10^8$ coupling constants. After an initial transient time $t_{\rm min}$, we have 
calculated a time-averaged value of the thermodynamic quantity of interest i.e. 
\begin{equation}
\label{timeaverage}
\langle Q\rangle_t ={1\over t-t_{\rm min}}\int_{t_{\rm min}}^{t} Q(t) dt
\end{equation}
and confirmed its convergence to a steady state value. An example of such simulations for the nodal powers $a_n(t)\equiv |\psi_n(t)|^2$ 
and the supermode power $|C_{\alpha}|^2$ are shown in Fig. \ref{fig2}a,b respectively. Failure to reach a steady state value indicated 
that the system did not reached the thermal equilibrium state. 

In the case of weak non-linearity the numerical results for $\left\langle |C_\alpha|^2\right\rangle$ have been compared against the 
theoretical predictions Eq. (\ref{C_alpha}). A good agreement between them serves as a confirmation that the system reached 
a thermal equilibrium state. A disagreement between the numerics and the theoretical predictions of Eq. (\ref{C_alpha}), indicates 
that the thermal equilibrium has not been reached. Instead, the system might have reached a metastable state as happens in the 
case of a spin-glass behavior \cite{N01,MPV87}. Given enough time (large relaxation times), of course, the system will reach the 
global free energy minimum. 

In all cases, the initial conditions were generated by considering the field amplitudes $\{|\psi_n|\}$, 
and the phases $\phi_n$ being random numbers in the intervals $\left[1-\delta,1+\delta\right]$ and $\left[-\pi,\pi\right]$ respectively. 
Out of a large number of $\{\psi_n\}$ configurations we have chosen only the ones that satisfy the energy and normalization constraints 
that define the specific state, see Eqs. (\ref{Hamiltonian},\ref{norm}) respectively. Finally, in all our simulations below we have used 
the normalization condition $A=\sum_n|\psi_n|^2 = N$ (i.e. $a=1$, see Eq. (\ref{aE})) associated with the total power.

\subsection{Equal Coupling Networks}\label{eqcoupling}

We start our analysis with the equal-coupling photonic network. In this case each node is coupled to all other nodes by a 
hopping amplitude of equal strength. The Hamiltonian that describes this system is Eq. (\ref{Hamiltonian}) with a connectivity 
matrix given by Eq. (\ref{conmat}) with $\sigma=0$. We have 
\begin{equation}
 {\cal H}\{\psi_l\}=-{J_0\over N}\sum_{l\neq j}\psi_l^*\psi_j+\frac12 \chi\sum_l|\psi_l|^4,
 \label{Ham_Jconst}
\end{equation}
where $l$ and $j$ run over all $N$ sites with the only constraint that $l\neq j$. 

In the absence of non-linearity the statistical mechanics of the system in Eq. (\ref{Ham_Jconst}) is trivial. The connectivity matrix 
(Eq. (\ref{conmat}) with $\sigma=0$) can be easily diagonalized. In this case the Hamiltonian Eq. (\ref{Ham_Jconst}) (with $\chi=0$)
has one non-degenerate eigenvector (supermode) with eigenfrequency $\varepsilon_1=-(N-1)J_0/N$ and $(N-1)$-fold degenerate 
eigenvectors with frequency $\varepsilon_{\alpha}=J_0/N (\alpha=2,\cdots,N)$. 
Therefore, in the large-$N$ limit, only the lowest non-degenerate mode contributes to the total energy $E$ and, since 
the latter quantity is required to be extensive, it is clear that a final fraction of the total optical power $A$ must reside in that mode. 
A simple calculation, based on relations Eq. (\ref{constrai}) and the expression Eq. (\ref{C_alpha}) yields the following result: Since the 
total optical power is $A=aN$ and the total energy is $E=hN$, then, in the large-$N$ limit, the resulting chemical 
potential and the temperature are
\begin{equation}
\label{mut}
 \mu=-J_0, \qquad T=J_0a+h.
\end{equation}
Since, obviously, $J_0a\geq|h|$ must hold, the linear model does not allow for either negative temperatures or for a transition: a finite 
fraction $(|h|/J_0a)$ of the total power $A$ is condensed into the lowest mode \cite{com3}. 

\begin{figure}
\includegraphics[width=\columnwidth]{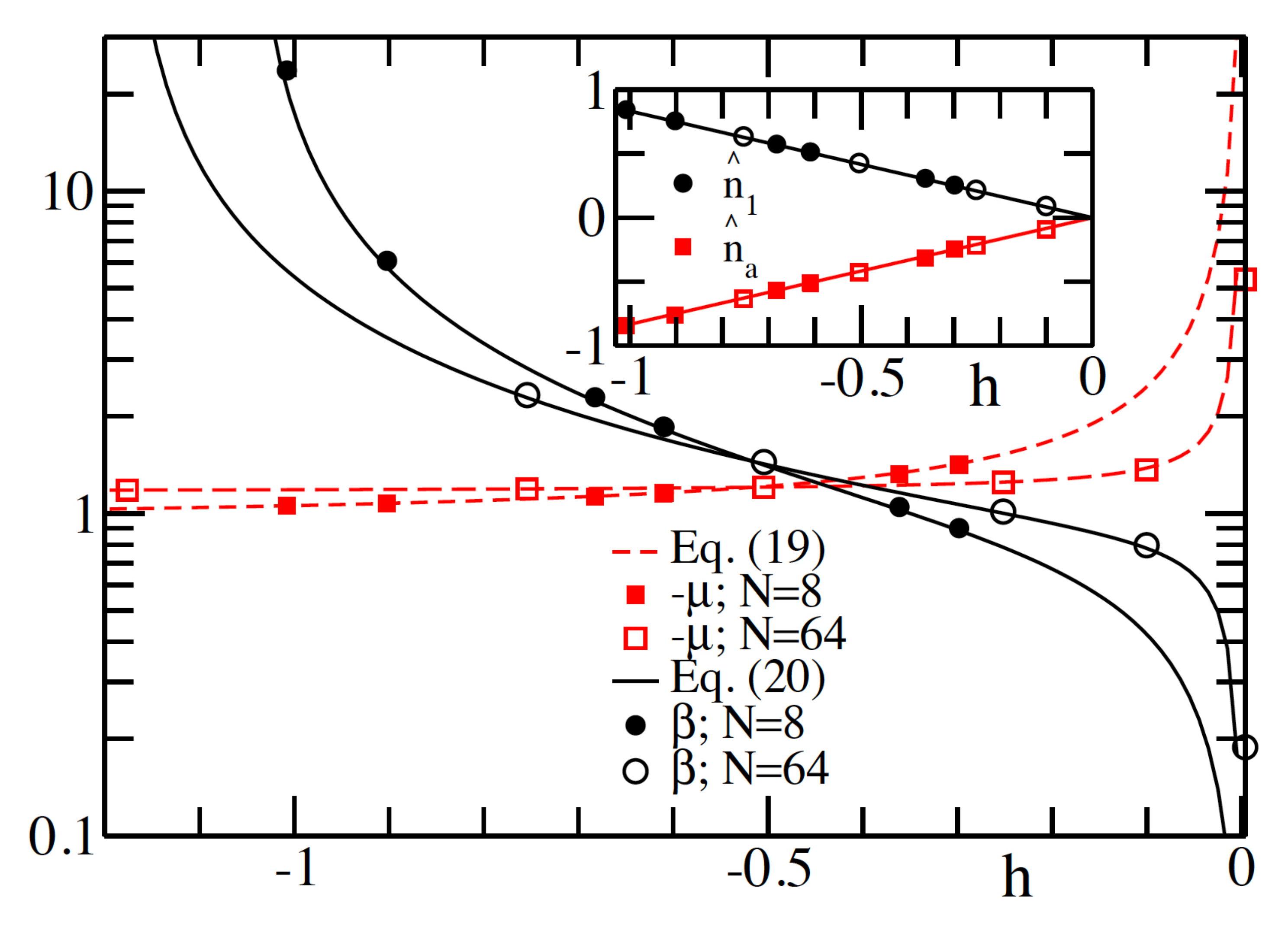}
\caption{(Color online) The numerical values of chemical potential $-\mu$ (red squares) and inverse temperature $\beta=1/T$ 
(black circles) versus the theoretical predictions Eqs. (\ref{refine_mu}) (red dashed line) and (\ref{refine_T}) (black lines) 
respectively. Filled symbols correspond to $N=8$, while open symbols correspond to $N=64$. (Inset) The numerically extracted 
optical powers (symbols) scaled as ${\hat n}_1$ and ${\hat n}_{\alpha}$ vs. the averaged 
energy density $h$. The rescaled ${\hat n}_1$ reaches the value zero at $h=h_{\rm max}$. The solid lines are the 
theoretical predictions of Eq. (\ref{C_alpha}) while the two type of symbols correspond to $N=8$ (filled symbols) and $N=64$ 
(open symbols). In all cases we considered an initial optical power $a=1$, coupling constant $J_0=1.2$ 
while the nonlinearity is weak $\chi=0.01$.}
\label{fig3}
\end{figure}

A refined analysis, where the finite size effects are taken into consideration, leads to the following exact expressions for the 
chemical potential 
\begin{equation}
\label{refine_mu}
\mu=-J_0\left[1-\left(a{J_0\over h}+2\right) {1\over N}+\left(a{J_0\over h}\right){1\over N^2}\right] 
\end{equation}
and the temperature
\begin{equation}
\label{refine_T}
T=J_0\left[\left(a+{h\over J_0}\right)-2\left({a\over N}\right)-\left({N-1\over N}\right) {J_0\over h} \left({a\over N}\right)^2\right]
\end{equation}
which, in the limit of $N\gg 1$, are nicely matching the results in Eq. (\ref{mut}). In Fig. \ref{fig3} we have compared these 
theoretical predictions with the values of $(\mu,T)$ that we have extracted from our numerical simulations with the Hamiltonian 
of Eq. (\ref{Ham_Jconst}), using two multimode photonic networks with $N=8$ and $64$ modes. The nice agreement indicates 
that these systems have reached a thermal equilibrium state. At the inset of the same figure we also report $\langle|C_{\alpha}|^2
\rangle_t$ (see Eq. (\ref{timeaverage})) by making use of the scaled variables ${\hat n}_\alpha$ 
\begin{eqnarray}
\label{scale}
{\hat n}_1\equiv \left(\langle|C_1|^2\rangle_t-a\right)/N=-h/J_0&\\
{\hat n}_{\alpha}\equiv \left(\langle|C_{\alpha}|^2\rangle_t-a\right) \times (N-1)/N=&h/J_0\nonumber
\end{eqnarray}
where in the second equation above $\alpha=2,\cdots,N$. The right hand side of Eqs. (\ref{scale}) has been evaluated 
using Eq. (\ref{C_alpha}) together with Eqs. (\ref{constrai}). It is important to point out that thermalization has been achieved 
even for the system with relatively small number of nodes $N=8$.

The presence of non-linearity changes the picture completely and provides us with an example of a (mean field) optical phase transition 
from an ordered to a disordered phase. In the former phase the amplitudes $\psi_l$ on different nodes are correlated while in the latter 
phase the nodes become essentially decoupled from each other. The ordered (disordered) phase corresponds to low (high) temperature, 
i.e. to small (large) internal energy $E$.

The ground state of the Hamiltonian (\ref{Ham_Jconst}) corresponds to a uniform field configuration $\{\psi_l\}=\sqrt{A/N}(1,1,\ldots ,1)$, 
where the normalization is such that the total optical power is $\sum_l|\psi_l|^2=A$. All the ``spins'' in this state point in the same direction. 
Note that the ground state is highly degenerated: one can rotate all the spins by an angle $\theta$, i.e. multiply the field by an overall phase 
factor $\exp(i\theta)$. The ground state energy is
\begin{equation}
\label{gstate}
E_{\rm min}=-J_0A+\frac12\chi A^2/N=\left(-J_0a+{1\over 2}\chi a^2\right)N 
\end{equation}
corresponding to $T=0$. In the opposite limit, $T\rightarrow\infty$, the kinetic energy (hopping) can be neglected and the system reduces 
to a set of uncoupled nonlinear oscillators. The probability density to find a node with optical power $|\psi|^2\equiv I$ is $p(I)\sim 
\exp(\beta\mu I-\frac12 \beta\chi I^2)$ and a simple calculation yields an average thermal energy equal to $\chi a^2$, in the $T\rightarrow
\infty$ limit. Thus, the maximal energy of the system (see Eq. (\ref{eq_hmax})) is $E_{\rm max}=\chi a^2N$ indicating that for fixed $a$ the 
thermalization occurs for $E_{min}<E<E_{max}$. One can, of course, endow the system with an energy larger than $E_{max}$, for instance, 
by putting all the norm $A$ on a single site (in which case the energy $E$ would become ``super-extensive'', proportional to $N^2$). We 
do not consider, however, initial preparation with $E>E_{\rm max}$ since we do not address the ``non-Gibbsian'' region in the $(E,A)$-plane 
(see previous discussion and also Refs. \cite{RCKG00,JR04}).

Next, we proceed with the calculation of the partition function Eq. (\ref{Zfunction}) which is performed using the saddle-point method. It is 
convenient to write the complex amplitudes $\psi_l=q_l+i\,p_l$ where $(q_l, p_l)$ is a pair of real variables. The Hamiltonian
\begin{equation}
 {\cal H}\{q_l, p_l\}=-\frac{J_0}{N}\sum_{l,j}\left(q_lq_j+p_lp_j\right)+\frac{\chi}{2}\sum_l \left(q_l^2+p_l^2\right)^2
\end{equation}
can be interpreted in terms of interacting two-component spins. We write $\sum_{l,j}q_lq_j=\left(\sum_lq_l\right)^2$ and use the identity
\begin{eqnarray}
 &&\exp\left[\frac{\beta J_0}{N}\left(\sum_l q_l\right)^2\right]=\\
 &&\sqrt{\frac{N\beta J_0}{\pi}}
 \int_{-\infty}^\infty dx\,\exp\left[\beta J_0\left(-N x^2+2\chi\sum_l q_l\right)\right]\nonumber
\end{eqnarray}
and similarly for $\sum_{l,j}p_lp_j$. The integrals over $q_l$, $p_l$ in the partition function now factorize into a product of integrals, 
each involving only variables for one node. Furthermore, in the large $N$-limit, the grand-canonical partition function ${\cal Z}$ is 
dominated by a saddle-point, which can be 
interpreted as the order parameter, i.e. the average field $\bar{\psi}$ (the magnetization in the statistical mechanics language). 
Actually, there is a whole family of saddle points, distinguished from each other by an overall phase. Choosing one saddle point of the 
family amounts to breaking the rotational symmetry in the spin space, obtaining a non-zero value for the order parameter. We chose 
$\bar{\psi}\equiv m$ to be real. Skipping all calculation details, we only give the final equation for $m$
\begin{equation}
 m=\frac{1}{2\beta J_0 F(m)}\frac{dF(m)}{dm}\equiv Q(m)
\label{16}
\end{equation}
where the function $F(m)$ is given by the integral $F(m)=2\pi\int_0^\infty r dr I_0(2\beta J_0mr)\exp\left(\beta\mu r^2-\frac{1} {2}\chi\beta 
r^4\right)$ and $I_0(x)$ is the modified Bessel function of order zero.

For small $m$, $Q(m)$ is a linear function of $m$, $Q(m)=s\cdot m$, and the slope $s$ determines whether Eq. (\ref{16}) has a nontrivial 
solution $m\neq 0$. Such solution exists only if $s>1$. The slope can be calculated using the small argument expansion $I_0(x)=1+
{1\over 4}x^2$, and it can be written as $s=\beta J_0K_3/K_1$ where an integral $K_n$ is defined as $K_n=\int_0^{\infty} dr r^n \exp(\beta
\mu r^2-{1\over 2}\chi \beta r^4)$. 

Let us, as an example, fix $\mu$ at the value zero and study $s$ as a function of the temperature $T=1/\beta$. For $\mu=0$ a simple 
expression for the slope $s$ is obtained, namely $s=J_0(2/\pi T\chi)^{1/2}$. The value $s=1$ yields the critical temperature $T_c=2J_0^2/\pi
\chi$. For temperature $T$ slightly below $T_c$ one finds, by keeping the term of order $m^3$ in the expansion of $Q(m)$, the standard 
mean field result $m\sim (T_c-T)^{1/2}$. When the temperature decreases further, towards $T=0$, the magnetization increases and 
approaches the maximal value, corresponding to the fully ordered ground state. Taking again $\mu=0$ as an example, one obtains from 
Eq. (\ref{16}) that, in the $\beta\rightarrow \infty$ limit, $m\rightarrow \sqrt{J_0/\chi}$. This result becomes transparent when written in terms 
of the optical power $A$. Indeed, at $T=0$ the total power resides in the (fully correlated) ground state so that the magnetization per site 
is $m=\sqrt{A/N}\equiv \sqrt{a}$. To connect $a$ to $\mu$ we have to use the expression Eq. (\ref{gstate}) for the ground state energy 
which yields $\mu={1\over N} \partial E/\partial a = -J_0 + \chi a$. For $\mu=0$, one obtains $m=\sqrt{a}=\sqrt{J_0/\chi}$.

\begin{figure}
\includegraphics[width=\columnwidth]{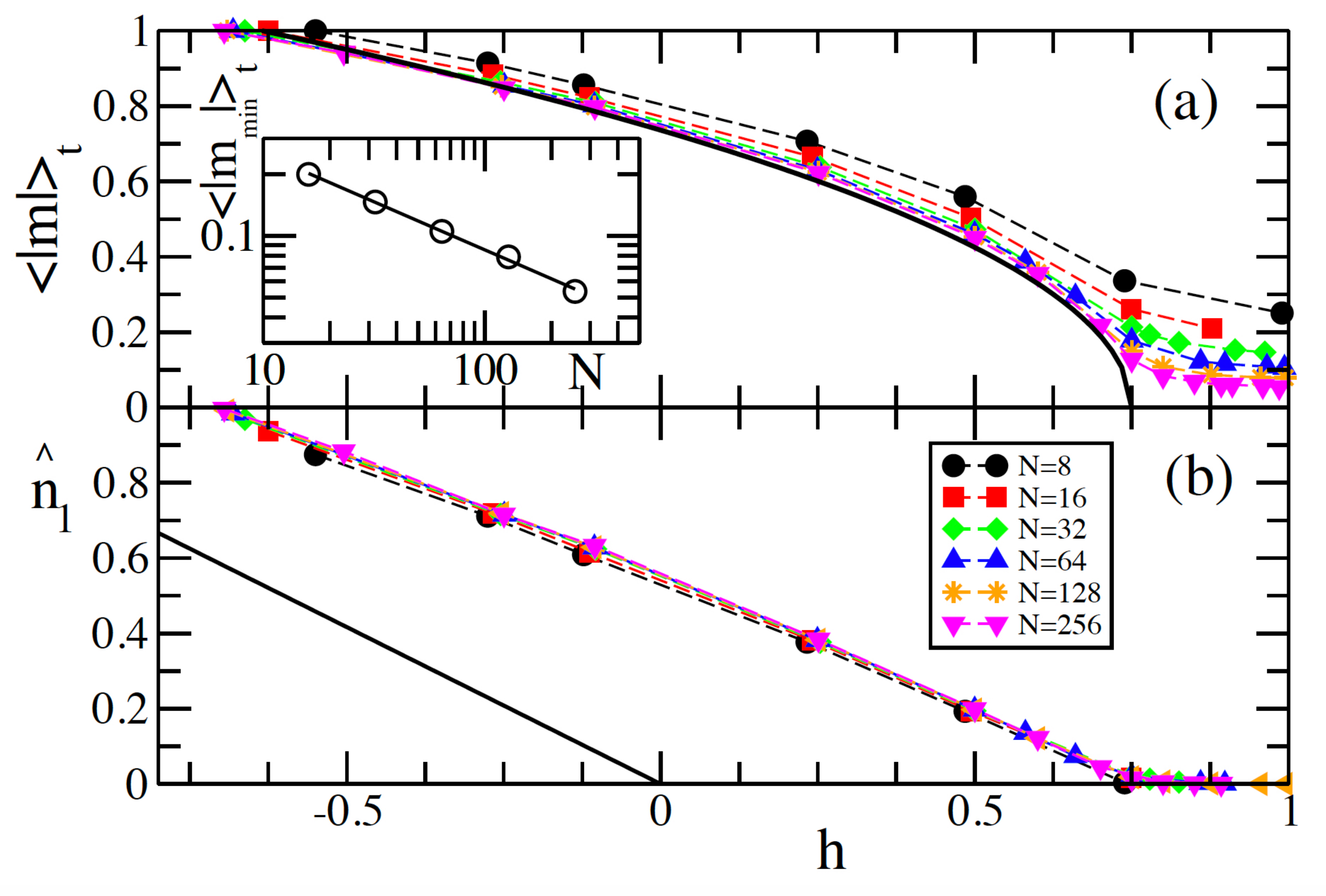}
\caption{(Color online) (a) The time-averaged magnetization versus the averaged energy density $h$. We have simulated various 
mulimode photonic networks described by Eq. (\ref{Ham_Jconst}) with $N=8,16,32,64,128,256$. The bold solid line is the best 
asymptotic ($N\rightarrow \infty$) fit indicating a square root singularity of the magnetization i.e. $\langle\left|m\right|\rangle_t\approx 
0.85\sqrt{0.75-h}$. Inset: The asymptotic value $\langle|m_{\rm min}|\rangle_t$ of time-averaged magnetization (circles), versus the 
number of nodes $N$. The solid line is the best fit to the numerical data and demonstrates a convergence to zero as $\langle|m_{\rm 
min}|\rangle_t\approx 0.75/\sqrt{N}$. (b) The time-averaged optical power of the ground state supermode for networks of different number 
of nodes $N$. We have used the same scaling variable ${\hat n}_1$ as in Fig. \ref{fig3}. The condensation transition, corresponding 
to ${\hat n}_1=0$, occurs for the same value of $h=h_c\approx 0.75$ as the one that signifies the transition from a ferromagnetic to 
a paramagnetic behavior in (a). In all cases we have considered an initial averaged optical power $a=1$, coupling constant $J_0=1.2$ 
while the nonlinearity is $\chi=1$. In this case, the maximum energy density is $h_{\rm max}=1$.
\label{fig4}}
\end{figure}

Thus, Eq. (\ref{16}) is well suited for studying $m$, as a function of $\beta$, for a fixed value of $\mu$. In experiment, however, one 
usually controls the optical power $A$, rather than the conjugate variable $\mu$. Calculating analytically $m$ as a function of $\beta$ 
for fixed $A$ is more involved than the above calculation for fixed $\mu$, and we do not attempt it in the present paper. Instead, in 
Fig. \ref{fig4} we present some numerical results, serving a double purpose: first to verify that the system, with the appropriate initial 
preparation, indeed thermalizes and then to study its properties in the thermal equilibrium state as a function of the averaged energy
density $h$. To this end, we evaluate numerically the time-averaged magnetization $\langle|m|\rangle_t$ (see Fig. \ref{fig4}a) defined 
as
\begin{equation}
\label{magnet}
\langle|m|\rangle_t=\langle\left|{1\over N}\sum_n\psi_n(t)\right|\rangle_t.
\end{equation}
In our numerics, we consider moderate values of the nonlinear parameter $\chi=1$ and coupling constant $J_0=1.2$. At the ground 
state $h=h_{\rm min}$  all ``spins'' have the same orientation. As a result, the magnetization acquires its maximum value $\langle|m|
\rangle_t=1$ indicating a ferromagnetic behavior. For higher $h$-values the magnetization decreases and at $h\rightarrow h_c\approx 
0.75$, which is smaller than $h_{\rm max}=1$, it acquires a constant value $\langle|m_{\rm min}|\rangle_t= {\cal O}(1/\sqrt{N})$ (see inset 
of Fig. \ref{fig4}a), associated with finite size effects. Further numerical analysis indicates that in the thermodynamic limit of $N\rightarrow
\infty$ the magnetization, as a function of $h$, approaches zero following a square-root behavior $\langle|m|\rangle_t\approx 0.85\sqrt{h_c-h}$ 
(see bold black line in Fig. \ref{fig4}a). Such behavior is characteristic of a second-order phase transition, from a ``ferromagnetic'' to 
a ``paramagnetic'' (optical) phase.

In Fig. \ref{fig4}b we show the numerical results for the time-averaged optical power in the ground state supermode $\langle|C_1|^2
\rangle_t$ versus the averaged energy density of the initial beam. In the simulations, we have considered the same photonic networks 
as above with $\chi=1$, $J_0=1.2$, and different numbers of nodes $N$. The numerical findings are reported using the scaled variable 
${\hat n}_1$, see Eq. (\ref{scale}). We find that for low averaged energy densities $h$, the optical power of the ground state is 
$\langle|C_1|^2\rangle_t\approx N$ indicating a condensate. As $h$ increases the condensate depletes and eventually at $h=h_c$, 
corresponding to the ferro-paramagnetic transition, it is completely destroyed i.e. ${\hat n}_1=0$. At the same figure we also report for 
comparison the theoretical value of ${\hat n}_1= -h/J_0$ (black solid line), applicable for the case of weak non-linearities, see Eq. 
(\ref{C_alpha}). We stress once more that the condensation transition analyzed above is due solely to the nonlinearity unlike the 
previous studied cases where the transition occurred already in the linear system \cite{AGMDP11,PR12}.

\subsection{Random Coupling} 

Next, we analyze the effect of disorder in the coupling constants of the photonic network i.e. $\sigma\neq 0$ in Eq. (\ref{conmat}). 
First, we consider the case of extreme disorder where $J_0=0$. In this case the coupling constants are entirely random, with equal 
probabilities to be positive or negative, i.e. the system is completely ``frustrated". Despite the vast literature on spin-networks, this 
model with complex, ``soft spins'' has not been studied up to now. Of course, certain analogies with the Sherrington-Kirkpatrick 
model can still be instructive. In the latter case, there is a transition from a paramagnetic phase to a spin-glass phase when the 
temperature (the energy $E$ of the system) decreases. One characteristic distinction between the two phases is that for the spin-
glass the thermalization time is much longer than for the paramagnet. Moreover, for large $N$ a spin-glass does not reach a full 
thermal equilibrium in any reasonable time, and the system gets stuck in one of the many metastable states.

\begin{figure}
\includegraphics[width=\columnwidth]{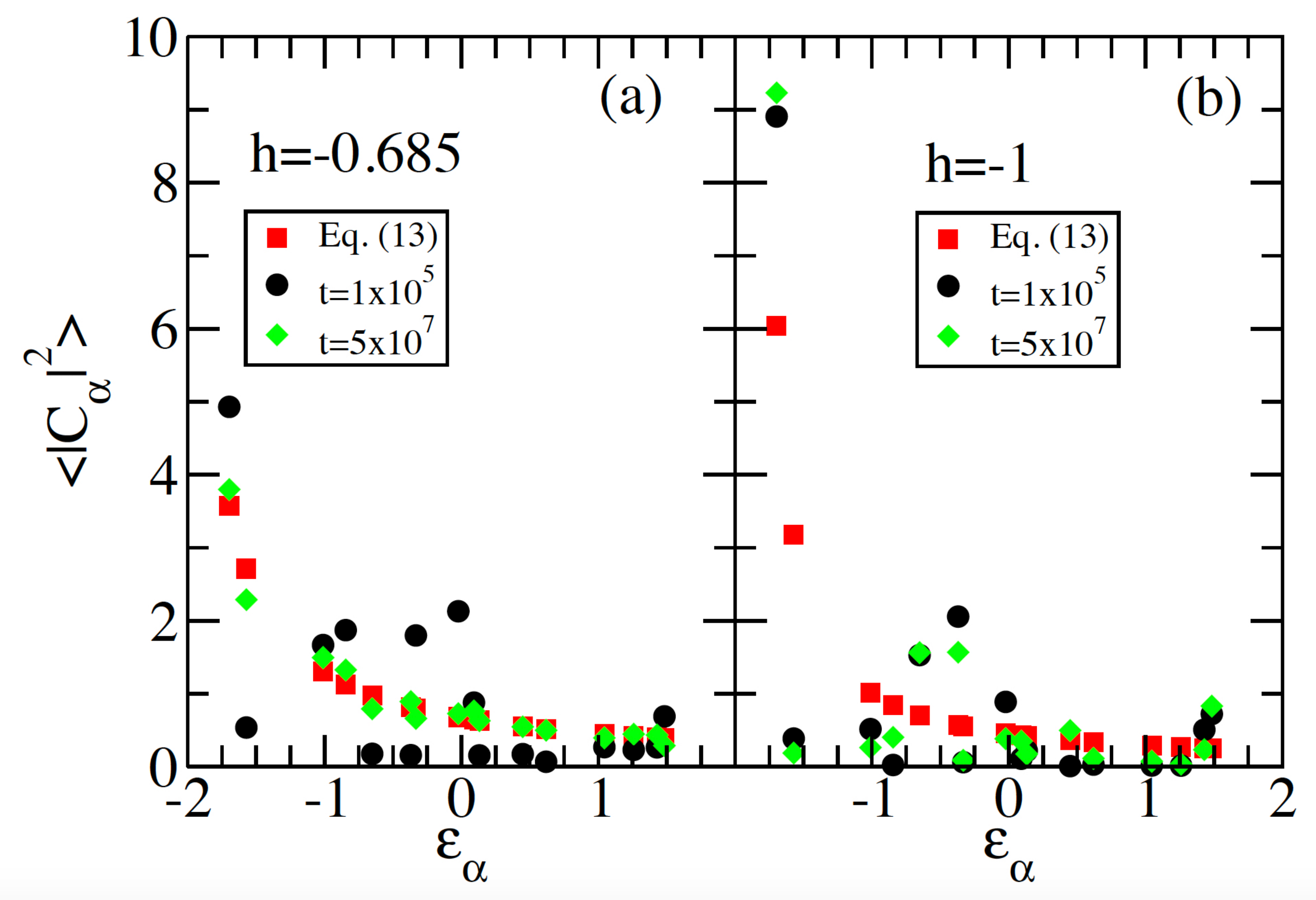}
\caption{(Color online) Comparison between the time-averaged supermode optical powers for a multimode photonic network whose 
dynamics is described by the Hamiltonian Eq. (\ref{Hamiltonian}) with random connectivity matrix 
Eq. (\ref{conmat}) with $J_0=0$ and $\sigma=1$, and the theoretical results (red squares) of Eq. (\ref{C_alpha}) for weak disorder. Black circles correspond 
to moderate time evolutions with $t=10^5$ while green diamonds correspond to larger time evolutions with $t=5\times 10^{7}$.
The initial state has (a) high averaged energy density per mode $h=-0.685$ (corresponding to high temperatures) or (b) low 
averaged energy density per mode $h=-1$ (corresponding to moderate/low temperatures). In both cases the average optical 
power is $a=1$. In these simulations, the number of modes is $N=16$ and the nonlinear parameter is $\chi=0.01$.
\label{fig5}}
\end{figure} 

Our simulations for the random coupling multimode photonic network are presented in Fig. \ref{fig5}. For a weak nonlinearity 
$\chi=0.01$, the equilibrium optical powers ${\bar n}_{\alpha}$ of the supermodes (of the linear problem) are given by 
Eq. (\ref{C_alpha}). In the simulation we evaluated the set $\{\left|C_{\alpha}(t)\right|^2\}$ as a function of time, extracted their 
time-average Eq. (\ref{timeaverage}), and use their comparison to Eq. (\ref{C_alpha}) as a criterion for thermalization. We find 
that for energy $h\approx-0.685$, see Fig. \ref{fig5}a, the system gets eventually close to thermal equilibrium at times $t\approx 
5\times 10^7$ (in units of standard deviation of the coupling elements). After this time the occupation numbers change only 
slightly. For higher energies (not shown) the thermalization time becomes shorter (for example for $h\approx-0.15$ the thermalization 
time for a network of $N=16$ nodes was $\propto 10^4$). On the other hand, for energy $h=-1$ (see Fig. \ref{fig5}b), the optical 
powers $\{\langle\left|C_{\alpha}\right|^2\rangle_t\}$ are far away from $\{{\bar n}_{\alpha}\}$ even after a fairly long time $t=5\times 
10^7$ and, moreover, they do not show any significant change with respect to the results extracted for shorter times $t=10^5$ 
(filled black circles in Fig. \ref{fig5}b). We have confirmed that the lack of thermalization (for any reasonable large time) is typical 
for other initial preparations (with the same $h$). This is a typical behavior of a spin-glass.

\begin{figure}
\includegraphics[width=\columnwidth]{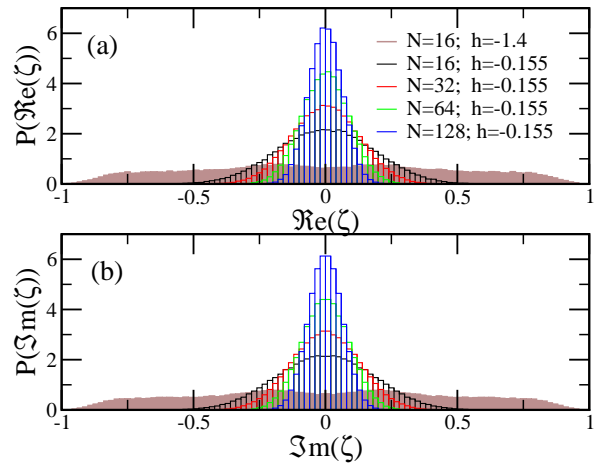}
\caption{(Color online) Distribution of the (a) real and the (b) imaginary part of the correlation function $\zeta(t)$. The ``empty'' histograms 
correspond to initial preparations with averaged energy density which is $h=-0.155$ (high energy regime). The brown ``filled'' histogram 
corresponds to a network with $N=16$ and initial preparations with averaged energy density which is $h=-1.4$ (low energy regime; 
$h_{\rm min}\approx\epsilon_1\approx-1.7)$. The number of modes are 
indicated by the color of the histogram in the inset of the figure. The average optical power was kept constant $a=1$ while $\chi=0.01$. 
In all these simulations the integration time was as long as $t=10^5$ and we have generated more than $32\times 10^4$ correlations 
$\zeta$ for the statistical processing.
\label{fig6}
}
\end{figure} 

Indeed, the most important signature of a spin-glass, from which the term itself was derived, is that at low temperatures the directions 
of spins at various sites get frozen in some random configuration (metastable state). For our ``optical spin glass'' such behavior implies 
that the average values of the complex amplitudes $\{\psi_l\}$, and in particular the phases $\{\theta_l\}$, form a random set. The 
``average'' here refers to the 
thermal statistical average, {\it for a fixed realization of the disorder}. One could expect that in a numerical simulation averaging over a statistical
ensemble can be replaced by the time average. However, due to a dynamic overall phase in the time evolution defined in Eq. (\ref{dif_schr}), 
the time-average of the phase, $\theta_l$, at any site $l$, amounts to zero. Therefore, in order to distinguish a spin glass from a 
paramagnet, we use the following criterion: Let us denote by superscripts $\alpha,\beta$ two initial preparations, with the same total 
energy $E$ and optical power $A$. Their time evolution is given by $\{\psi_l^{\alpha}(t)\}$ and $\{\psi_l^{\beta}(t)\}$ respectively. The quantity 
$\zeta(t)={1\over N}\sum_l\left( \psi_l^{\alpha}(t)\right)^*\psi_l^{\beta}(t)$ is a measure of the overlap between the two evolutions, at time $t$.
In the paramagnetic phase $\zeta(t)$ decreases with time, approaching zero, because (in the large $N$-limit) the two evolutions become
completely uncorrelated.

It is appropriate to consider many initial preparations, i.e. many $(\alpha,\beta)$-pairs, and treat the real and the imaginary parts of $\zeta(t)$ 
as statistical quantities with probability distribution ${\cal P}\left({\cal R}e(\zeta)\right)$ and ${\cal P}\left({\cal I}m(\zeta)\right)$. In the long 
time limit, and for large but finite $N$, these distributions are expected to have a manifestly different form in the two phases: For a paramagnet 
${\cal P}\left({\cal R}e(\zeta)\right)$ and ${\cal P}\left({\cal I}m(\zeta)\right)$ should be narrow distributions (with a width approaching zero when
$N\rightarrow \infty$), centered around $\zeta=0$. These expectations are confirmed by our numerical data which are reported in Figs. \ref{fig6}a,b 
for photonic multimode networks with initial preparations having high values of averaged energy density $h=-0.155$. An increase of the 
number of modes $N$ leads to narrower distributions around $\zeta=0$. When, on the other hand, we consider the same distributions for
a set of initial preparations with low value of $h=-1.4$, we observed an entirely different behavior for ${\cal P}\left({\cal R}e(\zeta)\right)$ and 
${\cal P}\left({\cal I}m(\zeta)\right)$ (see brown highlighted histogram in Figs. \ref{fig6}a,b). Namely, they become broad and almost flat, covering 
the whole allowed interval i.e. $-1<{\cal R}e(\zeta),{\cal I}m(\zeta)<1$. We stress that the above simulations were performed for a given realization 
of disorder and for the same energy $h=-1.4$. Only the initial preparations have been randomly chosen. We interpret the ``flatness'' of 
${\cal P}\left({\cal R}e(\zeta)\right),{\cal P}\left({\cal I}m(\zeta)\right)$ as a signature of many metastable states, typical of a spin-glass \cite{N01,MPV87}. 

\begin{figure}
\includegraphics[width=\columnwidth]{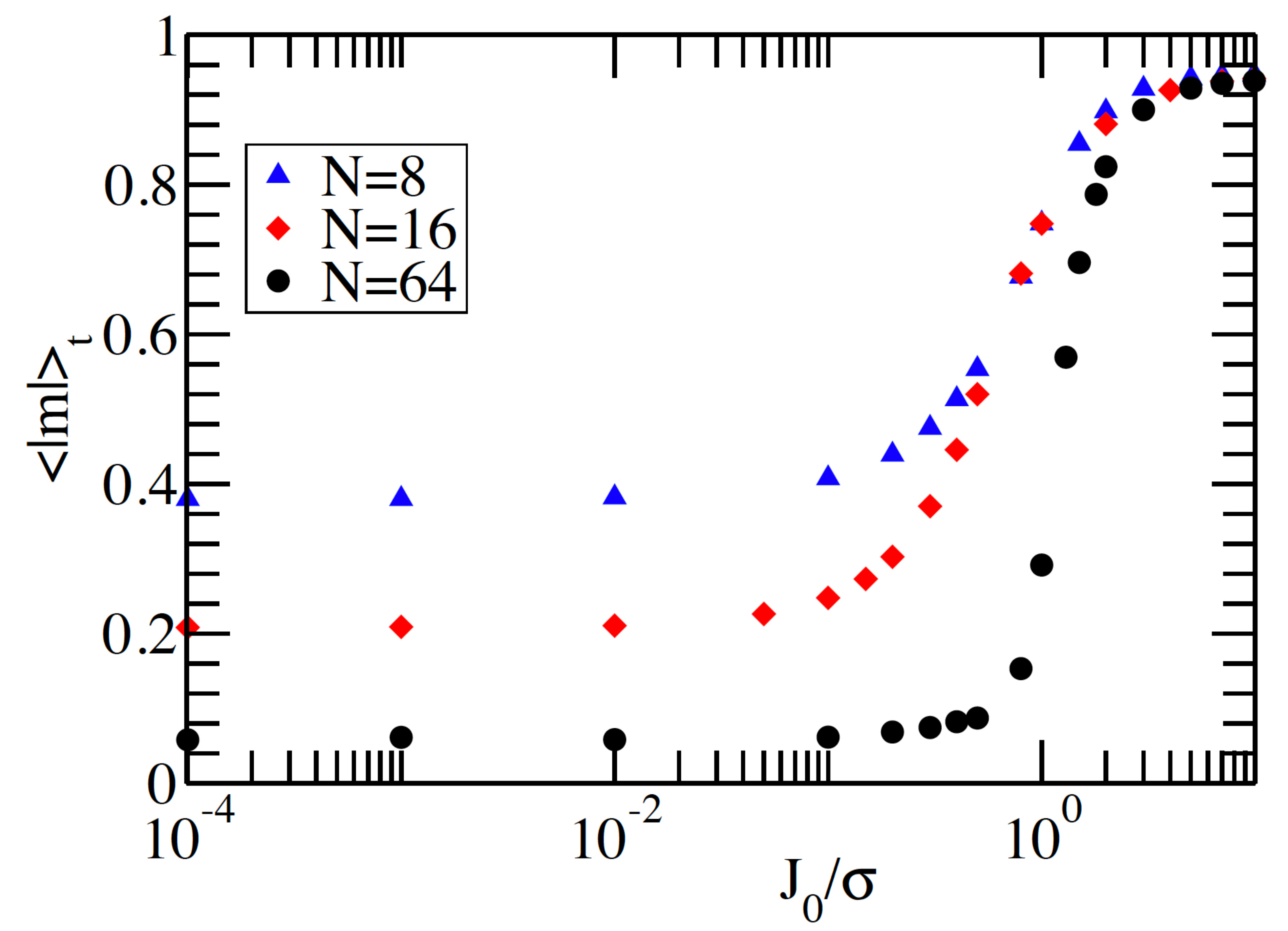}
\caption{(Color online) The time-averaged magnetization versus the control-parameter $x=J_0/\sigma$. The transition at $x\approx 1$ 
becomes more abrupt as $N$ increases. In these simulations, the initial preparation was taken to have averaged energy density $h\approx 
0.88 \epsilon_1$ (low energy regime) while the nonlinearity is $\chi=0.01$. 
\label{fig7}}
\end{figure}

It is natural to ask what happens to the network at low averaged energy densities $h$ when both terms in the connectivity matrix Eq. 
(\ref{conmat}) coexist, i.e. $J_0\neq0$ and $\sigma\neq 0$.  In Fig. \ref{fig7} we report the dependence of the magnetization $\langle
\left|m\right|\rangle_t$ versus the control parameter $x=J_0/\sigma$ and for $h\approx 0.88\epsilon_1$. In the simulations we keep 
$\sigma=1$ and change the magnitude of $J_0$. Following the same scheme as in section \ref{eqcoupling}, we break the rotational 
symmetry of the spin space by preparing the system at a real-valued configuration $\{\psi_n\}$. When $x=0$ the connectivity matrix 
has only random coupling elements (i.e $J_0=0$) ``forcing'' the network into the spin-glass phase. In this regime, the system evolves 
towards a metastable state with the ``spins'' pointing towards random directions. As a result, the magnetization is approaching zero as 
$\langle\left|m\right|\rangle_t\propto {\cal O}(1/N)$ in the limit of large $N$-values, see Fig. \ref{fig7}. In the other limiting case of 
$x\rightarrow\infty$ the randomness in the coupling elements are suppressed and 
the connectivity matrix is dominated by (essentially) constant couplings $J_0$. In this case, the network is in the ferromagnetic phase and 
the magnetization acquires a non-zero magnitude $\langle\left|m\right|\rangle_t\neq0$ which is dictated by the value of $h$ (e.g. for $h=
h_{\rm min}\approx -J_0$ the magnetization is $\langle\left|m\right|\rangle_t=1$). Our analysis (see Fig. \ref{fig7}) indicates that the transition 
from a spin-glass to a ferromagnetic phase occurs at $x\sim 1$. The transition becomes more abrupt, as expected from statistical mechanics, 
in the limit of large $N$-values.

\section{Conclusions}\label{concl}
We unveiled a connection between nonlinear photonic networks, consisting of many coupled single modes, and spin-networks. As opposed 
to standard spin models, our ``photonic spins'' are complex, soft (i.e. their size fluctuates), and their dynamics has two constants of motion: the 
total energy and the total optical power. This second conservation law is responsible for the appearance of novel optical phase transitions 
and the emergence of new forms of thermal photonic states. We have found that these transitions are controlled by the nature of the connectivity 
of the network (disorder or constant), and the amount of the averaged energy density of the initial optical preparation. Another important parameter 
is the strength of the non-linearity that controls the thermalization process. For strong non-linearities and constant couplings, we have discovered 
a ferro-paramagnetic phase transition as the averaged energy density $h$ of an initial preparation increases. This transition is associated with 
the destruction of a photonic condensate and its depletion to thermal states. In the other limiting case of random coupling constants the photonic 
network is driven from a paramagnetic to a spin-glass phase as $h$ decreases. Finally, we have shown that the same network, when prepared 
at low energies, undergoes another transition from a spin-glass to a ferromagnetic phase. The control parameter that drives this transition is the
degree of randomness (frustration) of the coupling constants between the photonic spins. Our results shed light on the ongoing effort of taming 
the flow of electromagnetic radiation in nonlinear multimode photonic networks. We also expect to sparkle the interest of the statistical physics 
community since the mathematical models that are typically used to describe light transport in multimode systems are falling outside the traditional 
spin-network framework.

{\it Acknowledgments --} We acknowledge useful discussions with D. Arovas, G. Bunin, J. Feinberg, L. Fern\'andez, and D. Podolsky. 
We also acknowledge the assistance of S. Suwunarrat with the preparation of Figure 1. This research was supported by an ONR via 
grant N00014-19-1-2480. (B.S) acknowledges the hospitality of WTICS lab at Wesleyan University where this work has been initiated.    


%
%


\begin{thebibliography}{10}
\bibitem{note1}Another kind of transition that can happens in $d=2$ is the Berezinskii-Kusterlitz-Thouless transition, studied in the 
context of photonic lattices in \cite{SPS11,SMF13}
\bibitem{com3}A transition, as well as, negative temperatures (inverted population) can only happen if the total energy is allowed to 
become non-extensive.
\end{thebibliography}
\end{document}